\documentclass[twocolumn,preprintnumbers,amsmath,amssymb,superscriptaddress,nofootinbib]{revtex4}

\usepackage{subfigure}
\usepackage{amsmath,amssymb}
\usepackage{graphicx}
\usepackage{rotating}
\usepackage{bm}
\usepackage{axodraw}
\usepackage{color}

\definecolor{blue}{rgb}{0,0,1}

\definecolor{green}{rgb}{0,1,0}

\newcommand{\rd}{\textrm{d}}
\newcommand{\ee}{\begin{equation}}
\newcommand{\eee}{\end{equation}}
\newcommand{\ea}{\begin{eqnarray}}
\newcommand{\eea}{\end{eqnarray}}

\definecolor{unterlegung}{rgb}{0.92,0.92,0.92}
\newlength{\importantlength}
\begin{document}

\author{Michael Doran}
\affiliation{Institut f\"ur Theoretische Physik, Philosophenweg 16,
  69120 Heidelberg} 
\author{Joerg Jaeckel}
\affiliation{Deutsches Elektronen Synchrotron, Notkestrasse 85,
22607 Hamburg}
\preprint{HD-THEP-06-09}
\preprint{DESY 06-079}
\title{What measurable zero point fluctuations can(not) tell us about dark energy.}

\begin{abstract}

We show that laboratory experiments cannot measure the absolute value of dark energy. 
All known experiments rely on electromagnetic interactions. They are thus insensitive to particles and
fields that interact only weakly with ordinary matter. 
In addition, Josephson junction experiments only measure differences in vacuum energy similar to Casimir
force measurements. Gravity, however, couples to the \emph{absolute} value. 
Finally we note that Casimir force measurements have tested zero point fluctuations up to energies of
$\sim 10\,\textrm{eV}$, well above the dark energy scale of $\sim 0.01\,\textrm{eV}$. Hence, the proposed 
cut-off in the fluctuation spectrum is ruled out experimentally. 
\end{abstract}
\maketitle

\section{Introduction}
Recently, a measurement of dark energy in 
the lab using noise in Josephson
junctions has been proposed \cite{Beck:2004fh}. While it would be wonderful to 
see dark energy in the lab, several groups argued that this will unfortunately
{\it not} be possible \cite{doran,Jetzer:2004vz,Mahajan:2006mw,Jetzer:2006pt}.
In the course of 
preparing this article, two groups \cite{Mahajan:2006mw,Jetzer:2006pt} 
published results similar to the
following arguments. We nevertheless would like to present our view of the topic,
hoping that it might help to clarify some aspects and settle the ongoing discussion.

Our reasoning is split in three parts: first, we argue that all devised measurements
of zero point fluctuations are based on the coupling to the electromagnetic field (Sect. \ref{electromagnetic}).
They would hence not measure all zero point fluctuations, but only those of particles 
coupled electromagnetically. Secondly, we show in Sect. \ref{josephsonjunction} 
that the fluctuation-dissipation theorem is invariant under a shift in vacuum energy. This
has already been observed by \cite{Jetzer:2006pt}. Thirdly,  in Sect. \ref{casimirresult}
we discuss experimental results for the Casimir force. Even if such experiments were to measure absolute
energies of zero point fluctuations (which they can't), the cut off in frequency needed to describe dark 
energy is not seen there. This argument  has been put forward 
earlier    \cite{doran}   and presented in more  detail recently \cite{Mahajan:2006mw}.

In a nutshell, our conclusion is: Although experiments studying Josephson junctions and the
Casimir effect might be interpreted as measuring zero point fluctuations, they cannot yield the absolute
amount of dark energy.

\section{Measurements need couplings to the electric field}\label{electromagnetic}
We are going to show that both Josephson junction type and Casimir effect experiments
are only sensitive to particles that couple to the $U(1)$ of the electromagnetic field.
To begin, consider the Casimir force between two ideal parallel 
plates  (see, e.g. \cite{Bordag:2001qi}), 
\begin{equation}
\label{casimir}
F(a)=-\frac{\partial E^{\textrm{ren}}_{0}(a)}{\partial a}, 
\quad E^{\textrm{ren}}_{0}(a)=-n_{\textrm{p}} \frac{c \hbar \pi^2}{720 a^3},
\end{equation}
where $n_{\textrm{p}}=2$ is the number of photon polarizations and $a$ is the distance between  plates.
The Casimir force \eqref{casimir} does not explicitly depend on the electromagnetic coupling.
Why then do only the fluctuations of
the electromagnetic field $n_{\textrm{p}}=2$ contribute? In other words, why is $n_{\rm p}$ not
replaced by the total number of degrees of freedom?

One might be tempted to attribute this to the suppression of contributions
of massive particles to the Casimir force $\sim \exp\left(-2 mca/\hbar\right)$.
Yet, modern experiments  (cf. also Sect. \ref{casimirresult}) probe 
scales $\frac{2\pi}{a}\sim 10\,\textrm{eV}$ where neutrinos are effectively massless.
No additional contributions from neutrinos have been found (as expected).
The solution is the very weak interaction of neutrinos with the plates. To
neutrinos, the plates are translucent and provide no boundary condition.
Photons, on the other hand, lead to a rearrangement of charges in the ideally
conducting plates such that the electric field vanishes. The charges enforce the
boundary conditions as a consequence of their coupling to the
electromagnetic field (the limit of ideally conducting plates
corresponds to an infinitely large coupling caused by the infinite
number of charges in the plates).  Accordingly, photons contribute to
the Casimir force as specified in \mbox{Eq. \eqref{casimir}}. 
From Eq. \eqref{casimir}, we also see that the Casimir force measures the
derivative of the zero point fluctuations and is thus insensitive to an overall shift $\Lambda$
in vacuum energy.

Turning to Josephson junctions, one measures the quantum mechanical
noise in voltage due to the noise of the electric current. 
 From this, the special role of
electromagnetically interacting particles is rather obvious. In
particular, Josephson junctions do not measure fluctuations of exotic weakly coupled
particles or even neutrinos.
 
\section{A closer look at the noise in Josephson junctions}\label{josephsonjunction}
The basis for the interpretation of the noise in Josephson junctions
as a measurement of vacuum fluctuations is the so called fluctuation
dissipation theorem,
\begin{equation}
\label{flucdiss}
\langle V^2\rangle=\frac{2}{\pi}\int^{\infty}_{0}R(\omega)
\left(\frac{1}{2}\hbar\omega+\frac{\hbar\omega}{\exp\left(\case{\hbar\omega}{kT}\right)-1}\right).
\end{equation}
Here $V$ is a ``force'' and $R(\omega)$ the ``resistance''.
The term 
\begin{equation}
\label{energy}
\frac{1}{2}\hbar\omega+\frac{\hbar\omega}{\exp\left(\case{\hbar\omega}{kT}\right)-1}
\end{equation}
closely resembles the vacuum energy of an harmonic
oscillator. Although this certainly  is an effect of zero point
fluctuations, it does not depend on the absolute value of the vacuum
energy:  shifting the Hamiltonian  by an arbitrary
constant $\Lambda$ (thereby changing the ``dark energy'' by this constant)
Eqs. \eqref{flucdiss} and \eqref{energy} remain unaffected. The 
constant $\Lambda$ will  not appear in the fluctuation-dissipation theorem.

To see this more clearly let us briefly review the  derivation
of Eq. \eqref{flucdiss} as given in \cite{callen}.  Starting point is
a Hamiltonian of the form
\begin{equation}
H=H_{0}(\cdots q_{k}\cdots p_{k}\cdots)+VQ(\cdots q_{k}\cdots p_{k}\cdots).
\end{equation}
$H_{0}$ is the unperturbed Hamiltonian and $V$ is a function of time that measures the magnitude of the perturbation while $Q$ depends only on the coordinates and momenta. In time dependent (quantum mechanical) perturbation theory one can now calculate the power absorbed by the system in an energy state $E_{n}$ when an outside force of the form $V=V_{0} \sin(\omega t)$ is applied,
\begin{eqnarray}
P(E_{n})\!\!&=&\!\!\frac{1}{2}\pi V^{2}_{0}\omega \big[
|\langle E_{n}+\hbar\omega |Q|E_{n}\rangle |^{2}\rho(E_{n}+\hbar\omega)
\\\nonumber
&&\quad\quad\quad\quad-|\langle E_{n}-\hbar\omega |Q|E_{n}\rangle |^{2}\rho(E_{n}-\hbar\omega)\big],
\end{eqnarray}
where $\rho(E)$ is the density of states with energy $E$. Weighting with a Boltzmann factor $f(E_{n})=\exp(-E/kT)/{\mathcal{Z}}$ and summing over all energy states one obtains the total power absorbed by the system,
\begin{eqnarray}
\label{power}
P_{tot}\!\!&=&\!\!\frac{1}{2}\pi V^{2}_{0}\omega \sum_{n}\big[
|\langle E_{n}+\hbar\omega |Q|E_{n}\rangle |^{2}\rho(E_{n}+\hbar\omega)
\\\nonumber
&&\quad\quad\quad\quad-|\langle E_{n}-\hbar\omega |Q|E_{n}\rangle |^{2}\rho(E_{n}-\hbar\omega)\big]f(E_{n}).
\end{eqnarray}
Let us now shift the Hamiltonian by a constant $\Lambda$ (we label
quantities in the ``new'' system with a tilde). For the 
matrix elements, the shift $\Lambda$ amounts to  relabeling
$E_{n}\rightarrow \tilde{E}_{n}=E_{n}+\Lambda$, i.e. the $n$-th state now has
energy $E_{n}+\Lambda$ but still the same wave function. 
Therefore, the matrix element remains the same. 
Likewise, the argument of the density of states is shifted by a constant $\Lambda$,
$\rho(E_{n})\rightarrow\tilde{\rho}(\tilde{E}_{n})=\tilde{\rho}(E_{n}+\Lambda)=\rho(E_{n})$. 

The only place where the energy enters explicitly is the Boltzmann factor
$f$. However, requiring proper normalization,
\begin{equation}
\sum_{n} f(E_{n})\rho(E_{n})=1=\sum_{n} f(\tilde{E}_{n})\tilde{\rho}(\tilde{E}_{n}),
\end{equation}
one easily finds using $\rho(E_{n}) = \tilde{\rho}(\tilde{E}_{n})$ that 
\begin{equation}
f(E_{n})=\tilde{f}(\tilde{E}_{n}).
\end{equation}
All in all, $P_{tot}$ is unaffected by the shift $\Lambda$ in energy.
Using the definition of impedance,
\begin{equation}
\label{force}
V=Z(\omega)\dot{Q}
\end{equation}
one has for the average dissipated power
\begin{equation}
\label{imppower}
P_{tot}=\frac{1}{2}V^{2}_{0}\frac{R(\omega)}{|Z(\omega)|^{2}}, \quad R(\omega)=Re(Z(\omega)).
\end{equation}
Using $\sum_{n}(\,\,)\rightarrow\int^{\infty}_{-\infty}(\,\,)\rho(E)dE$ we can replace the sum over states in Eq. \eqref{power} with an integral and compare to Eq. \eqref{imppower},
\begin{eqnarray}
\label{power_two}
\frac{R(\omega)}{|Z(\omega)|^{2}}\!\!&=&\!\!\pi\omega\int^{\infty}_{-\infty}dE\,\rho(E)f(E)
\\\nonumber
&&\quad\quad\quad\times\big[|\langle E+\hbar\omega |Q|E\rangle|^{2}\rho(E+\hbar\omega)
\\\nonumber
&&\quad\quad\quad\quad
-|\langle E-\hbar \omega|Q|E\rangle|^{2}\rho(E-\hbar\omega)\big].
\end{eqnarray}
Please note that although the integral goes from $-\infty$ to $\infty$ it effectively has a finite lower limit since the density
of states vanishes below the ground state energy of the system.

The next important step in the derivation of Eq. \eqref{flucdiss} is the calculation of $\langle V^2\rangle$.
Employing Eq. \eqref{force} it suffices to calculate $\langle \dot{Q}^2\rangle$.
One finds,
\begin{eqnarray}
\langle E_{n} |\dot{Q}^2| E_{n}\rangle =\sum_{m} (E_{n}-E_{m})^{2}|\langle E_{m}|Q|E_{n}\rangle|^{2}.
\end{eqnarray}
Replacing the sum by an integral and defining
\begin{equation}
\label{frequency}
\hbar\omega= | E_{n}-E_{m}|
\end{equation}
one obtains
\begin{multline}
\langle E_{n} |\dot{Q}^2| E_{n}\rangle =  \int^{\infty}_{0} \hbar \omega^2
\big[ |\langle E_{n}+\hbar \omega| Q|E_{n}\rangle \rho(E+\hbar\omega)
\\
\quad\quad+|\langle E_{n}-\hbar\omega|Q|E_{n}\rangle|^{2}\rho(E-\hbar\omega)\big] \rd{\omega},
\end{multline}
where the two parts originate from a splitting for $E_{n}>E_{m}$ and $E_{n}<E_{m}$.

Please note, that $\hbar\omega$ in \eqref{frequency} is a
\emph{difference} of energies. Again, a constant shift $\Lambda$ in the
Hamiltonian and therefore in the energy levels does not affect the
result.

Using Eq. \eqref{force} and integrating over all energy states weighted by a Boltzmann factor, one obtains
\begin{multline}
\label{forcesquared}
\langle V^{2}\rangle=\int^{\infty}_{0} |Z|^{2}\hbar\omega^{2}\bigg[\int^{\infty}_{-\infty}
\rho(E)f(E)
\\
\times\big[ |\langle E+\hbar\omega|Q|E\rangle|^{2}\rho(E+\hbar\omega)
\\
+|\langle E-\hbar\omega|Q|E\rangle|^{2}\rho(E-\hbar\omega)\big]dE\bigg]d\omega.
\end{multline}
Following \cite{callen}, we denote the integrals over $E$ in Equations \eqref{power_two} and \eqref{forcesquared} by
\begin{multline}
C_{\pm} = \int_{-\infty}^\infty f(E) \rho(E) \biggl \{ 
|\langle E+\hbar\omega|Q|E\rangle|^{2} \rho(E+\hbar\omega)
\\ \pm
|\langle E-\hbar\omega|Q|E\rangle|^{2}\rho(E-\hbar\omega)
\biggr \} \rd E
\end{multline}
One can 
shift $E \to E + \hbar \omega$ in the second term of $C_{\pm}$ yielding
\begin{multline}
C_{\pm} =  \int_{-\infty}^\infty \rho(E) \rho(E+\hbar\omega)\,|\langle E+\hbar\omega|Q|E\rangle|^{2} \\
 \times f(E)  \left \{1 \pm \frac{ f(E+\hbar\omega) } {  f(E) }   \right\} \rd E.
\end{multline}
As $ f(E+\hbar\omega)  / f(E) = \exp(-\hbar \omega / kT)$, one therefore gets
\ee
C_{\pm} = \left (1 \pm \exp\left(-\frac{\hbar \omega}{k T}\right)\right) C(\omega),
\eee
where
\ee
C(\omega) = \int_{-\infty}^\infty f(E) \rho(E)\,\rho(E+\hbar\omega)\, |\langle E+\hbar\omega|Q|E\rangle|^{2} \,\rd E.
\eee
Equations  \eqref{power} and \eqref{forcesquared}  may therefore be written as
\ee\label{R_final}
R(\omega) = \pi \omega |Z(\omega)|^2  \left (1 -   \exp\left(-\frac{\hbar \omega}{k T}\right) \right)  C(\omega)
\eee
and
\ee\label{forcesquared_final}
\langle V^2 \rangle = \int_0^\infty \rd \omega |Z(\omega)|^2 \hbar \omega^2 \left (1 +  \exp\left(-\frac{\hbar \omega}{k T}\right) \right)  C(\omega)
\eee
Finally, combining Eqs. \eqref{R_final} and \eqref{forcesquared_final} one arrives at \eqref{flucdiss}.
Hence, the derivation of the fluctuation-dissipation theorem of Callen and Welton is unaffected by a 
constant shift $\Lambda$ of the Hamiltonian. In particular, it does not appear in Equation \eqref{energy}.

To conclude this section let us  in addition remark that Eq. \eqref{flucdiss} is a
statement about quantum mechanical and thermal fluctuations in an
arbitrary system. This could, e.g. be  fluctuations of a
quantum harmonic oscillator in a thermal bath. These fluctuations are
not necessarily in one to one correspondence to the fluctuations in
 fundamental quantum fields. For example, the particle in the
harmonic oscillator could itself be a bound state.
Moreover, the common factor Eq. \eqref{energy} that 
is interpreted as the effect of vacuum fluctuations is
essentially the average energy of one harmonic oscillator of frequency $\omega$. However, the total vacuum energy depends on the density of states $\rho(E)$, which will be different from the true vacuum in the solid state setup of the Josephson junction\footnote{This poses a severe problem for any argument concerning dark energy in which Josephson junctions play a special role, e.g. \cite{Beck:2006pv}.}. 

\section{Vacuum fluctuations have already been measured beyond 10eV}\label{casimirresult}
In recent years there has been significant progress in the measurement
and calculation of the Casimir force. Experiments have reached 
distances $d$ below $d\sim 100\,\textrm{nm}$ (see, e.g.,  \cite{harris}).

Separations of $d\sim 100\,\textrm{nm}$ correspond to energy scales of
 $\frac{2\pi}{d}\sim 10\,\textrm{eV}$. Hence, measurements of the Casimir force already
probed  photon zero point fluctuations with
energies in excess of $10\,\textrm{eV}$. 
The cosmologically inferred scale of dark energy on the other hand is only 
of the order of $(\case{8\pi^2\,c^3}{\hbar} \rho_{\textrm{DE}})^{\frac{1}{4}}\sim 0.01\,\textrm{eV}$ ($\rho_{\textrm{DE}}$ is the current energy density of dark energy). This is much smaller than the scales
already probed by Casimir force experiments. The hypotheses that 
a cut off in the fluctuation spectrum might be responsible for dark 
energy has thus been excluded experimentally: the Casimir measurements
see no such cut off. 

Please note that these energies are much larger
than those probed so far in the measurements of  noise in Josephson
junctions where  frequencies up to $6 \times 10^{11} \,{\rm Hz}$ corresponding to
$2.5 \times 10^{-3}\,\textrm{eV}$ have been reached. When Josephson junction
experiments reach $\sim 0.01\,\textrm{eV}$, they should hence see no cut off,
in agreement with Casimir force experiments.

\section{Conclusions}\label{conclusions}
In this brief note we tried to clarify some aspects of the
question what measurements of zero point fluctuations can tell us
about dark energy. We argued that in all known experiments, we only measure 
fluctuations of fields interacting electromagnetically.
These experiments are insensitive to fluctuations of weakly interacting 
particles.  For instance, a mechanism that adds a very
weakly interacting fermion to cancel the vacuum energy contributed by
the photon would not be seen in those experiments.
Josephson junction experiments will not measure the absolute value of 
dark energy, because the noise is independent of a constant shift of the Hamiltonian.
From Casimir force measurements,
we know that  the photon has zero
point fluctuations with energies above $10\,\textrm{eV}$,
much above the dark energy scale of $\sim0.01\,\textrm{eV}$. Hence, some
strange cutoff  that only affects zero point fluctuations and
thereby restricts  dark energy is ruled out\footnote{A possible
loop hole is that such a cutoff affects only fluctuations that truly
do not couple (aside from gravity). But those, again wouldn't be
measurable.}.


\newpage
\begin{thebibliography}{99}
\bibitem{Beck:2004fh}
  C.~Beck and M.~C.~Mackey,
  Phys.\ Lett.\ B {\bf 605} (2005) 295
  [astro-ph/0406504].

\bibitem{doran}
Doran, M. in: Chown, M. A., 
New Scientist {\bf 183} 2455 (2004) 11.

\bibitem{Jetzer:2004vz}
  P.~Jetzer and N.~Straumann,
  Phys.\ Lett.\ B {\bf 606} (2005) 77
  [astro-ph/0411034].

\bibitem{Jetzer:2006pt}
  P.~Jetzer and N.~Straumann,
  astro-ph/0604522.

\bibitem{Mahajan:2006mw}
  G.~Mahajan, S.~Sarkar and T.~Padmanabhan,
  astro-ph/0604265.

\bibitem{Beck:2006cv}
  C.~Beck and M.~C.~Mackey,
  astro-ph/0603397.

\bibitem{Bordag:2001qi}
  M.~Bordag, U.~Mohideen and V.~M.~Mostepanenko,
  Phys.\ Rept.\  {\bf 353} (2001) 1
  [quant-ph/0106045].

\bibitem{callen} H. Callen and T. Welton, Phys. Rev. {\bf 83}, 34
(1951).

\bibitem{harris}
B.W.~Harris, F.~Chen, U.~Mohideen, 
Phys. Rev. A 62 (2000) 052109.

\bibitem{Beck:2006pv}
C.~Beck and M.~C.~Mackey,
astro-ph/0605418.
        
\end{thebibliography}
\end{document}